\documentclass[12pt,a4paper]{article}
			     \input epsf
\begin{document}
\title{\hfill\raise 20pt\hbox{\small DIAS-STP-05-06}\\
N=2 Supersymmetric Yang-Mills and the Quantum Hall Effect}
\author{Brian P. Dolan\\
Dept. of Mathematical Physics, \\
National University of Ireland, Maynooth\\
and \\
Dublin Institute for Advanced Studies,\\
10, Burlington Rd., Dublin, Ireland\\
{\tt bdolan@thphys.nuim.ie}\\}

\maketitle

\begin{abstract}
It is argued that there are strong similarities between the infra-red
physics of N=2 supersymmetric Yang-Mills and that of the quantum
Hall effect, both systems exhibit a hierarchy of vacua with a
sub-group of the
modular group mapping between them.
The coupling flow for pure $SU(2)$ $N=2$ supersymmetric Yang-Mills
in 4-dimensions is re-examined and an earlier suggestion in the 
literature, that was singular at strong coupling,
is modified to a form that is well behaved at both weak and strong coupling
and describes the crossover in an analytic fashion.  
Similarities between the phase diagram and the flow
of SUSY Yang-Mills and that of the quantum Hall effect are then described,
with the Hall conductivity in the latter
playing the role of the $\theta$-parameter in the former. 
Hall plateaux, with odd denominator filling fractions,
are analogous to fixed points at strong coupling in N=2 SUSY Yang-Mills, 
where the massless degrees of freedom carry an odd monopole charge.


\end{abstract}

\section{Introduction}

In this paper arguments will be given that there there are
strong similarities in the infra-red physics of $N=2$ supersymmetric
Yang-Mills and that of the quantum Hall effect. 
A common feature of these two systems
is the emergence of modular symmetry (strictly a sub-group of the
modular group) in the infra-red regime.  That the
modular group may be relevant to the quantum Hall effect was first
suggested in \cite{ShapereWilczek}, though the correct subgroup
was only found later \cite{LR1}.

In the infra-red limit both theories can be parameterised by 
a complex parameter whose real part is the co-efficient of a topological
term in the effective action describing the infra-red physics
and whose imaginary part is essentially the co-efficient of the kinetic 
term, which must be positive.  
In $N=2$ SUSY the complex parameter is $\tau={\theta\over 2\pi}+{4\pi i\over g^2}$ with $\theta$ the QCD vacuum parameter and $g^2$ the Yang-Mills
coupling (in units with $\hbar=c=1$), in the quantum Hall 
effect the complex parameter is
$\sigma=\sigma_{xy} +i\sigma_{xx}$ where $\sigma_{xy}$ is the
Hall conductivity and $\sigma_{xx}$ the Ohmic conductivity.
In both cases the complex parameter is constrained to lie in the
upper-half complex plane for stability reasons.

In the low energy effective action for supersymmetric Yang-Mills
$\theta$ is proportional to the co-efficient of the topological 
term $F\wedge F$ 
and $1/g^2$ is proportional to
the co-efficient of the kinetic $F^2$ term \cite{SW}.
In the low energy effective action for the electromagnetic field in
quantum Hall effect
$\sigma_{xy}$ is proportional to the co-efficient of the Chern-Simons 
term and $\sigma_{xx}$
is related to the co-efficient of the effective kinetic term  \cite{KLZ}
(in a conducting medium the dielectric constant is the co-efficient of
$\vec{E}.\vec{E}$ in the long wavelength effective action and
has a simple  pole at zero frequency, the Ohmic 
conductivity is the residue of the pole).

Both systems have a hierarchy of phases:
different strong coupling vacua of SUSY Yang-Mills on the one hand 
and different quantum Hall plateaux on the other.    
These different phases are mapped into  each other by the
action of a sub-group of the modular group.

There are also remarkable similarities between the scaling flow
of the complex couplings in both systems and the main focus of this
paper is to examine this relation.  Mathematically the link is
the modular group.

In section 2 the scaling flow for $N=2$
SUSY Yang-Mills without matter is discussed and a flow 
of the effective complex coupling as the Higgs VEV is varied
is constructed which is
compatible with $\Gamma_0(2)$ symmetry and 
reduces to the Callan-Symanzik $\beta$-function near the fixed points.  
The flow is
a modification of the flow described in \cite{MN}, that
had singularities at strong coupling, and these singularities
are avoided in the flow proposed here.
Striking
similarities with the temperature flow of the conductivities of the
quantum Hall effect are described in section 3, 
together with other parallels between these two systems.

\section{$N=2$ SUSY $SU(2)$ Yang-Mills}

After the seminal results of Seiberg and Witten \cite{SW},
in which the low energy effective action for $N=2$ supersymmetric Yang-Mills
theory in 4-dimensions was explicitly constructed, 
scaling functions were derived explicitly for particular gauge groups 
and matter content in \cite{MN}.
The case of pure $SU(2)$ Yang-Mills was taken further in 
\cite{RDBMLL}.  The scaling function can be  constructed in terms 
the complex coupling $\tau={\theta\over 2\pi}+{4\pi i\over g^2}$
and the vacuum expectation value of the Higgs field, or better
the gauge invariant mass scale, $u=tr<\varphi^2>$, where $\varphi$ 
is the Higgs field in the adjoint of $SU(2)$.
A non-zero vacuum expectation value for $\varphi$ gives the
$W^\pm$-bosons a mass and breaks the gauge symmetry down to $U(1)$.
The scaling function introduced in \cite{MN} was essentially
the logarithmic derivative
of the Seiberg-Witten low energy effective
coupling $\tau(u)$ with respect to $u$,
\begin{equation}
\Sigma_0(\tau)=-u{d\tau\over du}=
{1\over 2\pi i}\left({1\over\vartheta^4_3(\tau)}+{1\over\vartheta^4_4(\tau)}\right),
\label{betazero}
\end{equation}
where $\vartheta_3(\tau)=\sum_{n=-\infty}^\infty e^{i\pi n^2\tau}$
and $\vartheta_4(\tau)=\sum_{n=-\infty}^\infty (-1)^ne^{i\pi n^2\tau}$
are Jacobi $\vartheta$-functions.
In the definitions of \cite{Koblitz} $\Sigma_0(\tau)$ is a modular function
of the group $\Gamma_0(2)$, of weight $-2$.\footnote{$\Gamma_0(2)$ is the
sub-group of the full modular group $\Gamma(1)$ consisting
of matrices $\gamma=\pmatrix{a&b\cr c&d\cr}\in\Gamma(1)$ with 
$a$, $b$, $c$ and $d$ integers, $ad-bc=1$ and $c$ 
restricted to be even.}

Although (\ref{betazero}) does take into account all 
non-perturbative effects, it is only valid in the weak-coupling regime
and was criticized in \cite{K1,K2}
since it is unphysical at strong coupling, giving a singularity
at $\tau=0$.  $\Sigma_0$ also has an attractive 
fixed point at $u=0$ ($\tau={1+i\over 2}$ and its images under $\Gamma_0(2)$)
which arises because 
the flow is defined to be
radially inwards towards the origin in the $u$-plane. 
We can change this and put the attractive fixed point anywhere in
the finite $u$-plane, without affecting the behaviour at infinity,
by defining a meromorphic scaling function
\begin{equation}
\Sigma_{u_0}(\tau)=-(u-u_0){d\tau\over du}.
\label{betace}
\end{equation}
From (\ref{betazero}) it is straightforward to express (\ref{betace})
as an explicit function of $\tau$ using elementary properties
of $\vartheta$-functions (see {\it e.g.} \cite{WW}). 
The effective coupling can be written in terms of elliptic
integrals \cite{GH}
\begin{equation}
\tau = i{K'(k)\over K(k)} +2n
\end{equation}
where $K(k)=\int_0^{2\pi}{1\over\sqrt{1-k^2\cos^2\phi}}d\phi$ is the complete
elliptic integral of the first kind, $k^2:={2\Lambda^2 \over \Lambda^2+u}$
with $\Lambda^2$ the QCD scale, 
$K'(k)=K(k')$ with
$k'^2=1-k^2$ and $n$ is an integer
($\tau$ is not uniquely determined by $k$, rather 
$e^{i\pi\tau} = e^{-\pi {K'\over K}}$).
In terms of $K(k)$ the $\theta$-functions are
\begin{equation}
\vartheta_3(\tau)=\sqrt{2K(k)\over\pi}\qquad \hbox{and} \qquad
\vartheta_4(\tau)=\sqrt{2k'K(k)\over\pi}.
\label{thetaK}
\end{equation}
Since $k^2=2\Lambda^2/(u+\Lambda^2)$ we obtain
\begin{equation}
{\vartheta_4^4(\tau)\over\vartheta_3^4(\tau)}=1-k^2=
{u-\Lambda^2\over u+\Lambda^2}
\qquad\Rightarrow\qquad
{u\over\Lambda^2}={\vartheta_3^4 + \vartheta_4^4\over \vartheta_3^4 - \vartheta_4^4}.
\end{equation}
Combining this with (\ref{betazero}) gives
\begin{equation}
\Sigma_{\tilde u_0}(\tau)=-(u-u_0){d\tau\over du}={1\over 2\pi i}\left(
{1-\tilde u_0\over\vartheta_4^4} + {1+\tilde u_0\over\vartheta_3^4}
\right)
\label{betac}
\end{equation}
with $\tilde u_0=u_0/\Lambda^2$.

The shift $\tau\rightarrow \tau
+1$ is equivalent to $u_0\rightarrow -u_0$ (this follows from the 
property of Jacobi $\vartheta$-functions,
$\vartheta_3(\tau+1)=\vartheta_4(\tau)$)
which in turn is a manifestation of the
the $Z_2$ action on the $u$-plane familiar from \cite{SW}.
Since (\ref{betac}) is not invariant under this shift when $\tilde u_0\ne 0$
it is not
a modular function of $\Gamma_0(2)$, rather it is a modular function
of weight -2 for the smaller group $\Gamma(2)$.\footnote{$\Gamma(2)$ 
consists
of matrices $\gamma=\pmatrix{a&b\cr c&d\cr}\in\Gamma(1)$ with 
$a$, $b$, $c$ and $d$ integers, $ad-bc=1$ and both $b$ and $c$ even.}
The singularity at $\tau=0$ present in $\Sigma_0(\tau)$ can be avoided by
choosing $\tilde u_0=1$, while that at $\tau=1$ is avoided by
using $\tilde u_0=-1$.
In either case the flow generated by $\Sigma_{\pm 1}$ runs along the 
semi-circle spanning $\tau=1$ to $\tau=0$,
this semi-circle is the straight-line segment in the $u$-plane
running between $u=+\Lambda^2$ and $u=-\Lambda^2$ and all its images under 
$\Gamma(2)$ are also semi-circles in the upper-half $\tau$-plane.  

Consider $\tilde u_0=+1$ (a similar analysis applies to $\tilde u_0=-1$),
\begin{equation}
\Sigma_{+1}(\tau)={1\over \pi i} {1\over\vartheta^4_3(\tau)}.
\label{betaone}
\end{equation}

The resulting flow in the $\tau$-plane shown in figure 1 and
is symmetric under translations in $\tau$
by 2, $\tau\rightarrow \tau \pm 2$.  
Any starting value of $\tau$ at weak coupling 
$u=\infty$ with $-2\pi<\theta < 2\pi$ 
is driven to $\theta=0$ at strong coupling
so consider the positive 
imaginary axis in the $\tau$-plane $\theta=0$, where $\tau=i{4\pi\over g^2}$.
In figure 2 the function $\Sigma_{+1}$ is plotted as a function of
$Im(\tau)={4\pi\over g^2}$ along the positive imaginary axis, $\theta=0$.  
Note the constant value $-1/\pi$ at small $g$,
consistent with asymptotic freedom behaviour 
$u{d g\over du}\approx -{g^3\over 8\pi^2}$.

For large $g$, $\Sigma_{+1}\approx {i\tau^2\over \pi}$ 
is perfectly well behaved and in terms of the dual coupling $g_D$, defined by 
$\tau_D=-{1\over \tau}$ so
$\tau_D=i{4\pi\over g_D^2}$ when $\theta=0$, 
one finds
$-(u-\Lambda^2) {d g_D\over du}\approx -{g_D^3\over 8\pi^2}$ in agreement with the expectations for the dual theory \cite{SW}.
$\Sigma_{+1}$ for the theory and its dual are mapped
onto each other using the identity 
$\vartheta_3(-1/\tau)=\sqrt{-i\tau}\vartheta_3(\tau)$ giving 
\begin{equation}
\Sigma_{+1}(\tau_D)={i\over\pi} {1\over\vartheta^4_3(\tau_D)}.
\label{betaplusoneD}
\end{equation}
This therefore is a well behaved scaling function which is defined
for both the the theory and its dual and gives the crossover
between weak and strong coupling.
It was suggested in \cite{K1} that the zero of $\Sigma_0$ in equation
(\ref{betazero}) at $u=0$
was spurious and that a better scaling function
would be 
\begin{equation}
\tilde\Sigma(\tau)=G(\tau)\Sigma_0(\tau)
\end{equation}
where $G(\tau)$ is a renormalisation factor with a pole at $u=0$
which accounts for the zero in $\Sigma_0$.
We see here that
$\Sigma_{+1}=\left({u-\Lambda^2\over u}\right)\Sigma_0$
is indeed related to $\Sigma_0$ by a simple pole in the $u$ variable
at $u=0$.

Near $u\approx\infty$ and  $u\approx +\Lambda^2$ the function
$\Sigma_{+1}$ 
behaves like a Callan-Symanzik $\beta$-function and this is the
physical difference from the scaling function considered in \cite{MN}. 
At weak coupling, where $u\approx\infty$, 
the gluinos have a mass $M$ proportional to the
the VEV of the Higgs, $<\varphi>=a$, and $u\approx a^2/2$ so $\Sigma_{+1}$
gives
\begin{equation}
\Sigma_{+1}\approx -u{d\tau\over du}\approx -a^2{d\tau\over da^2}
\approx -M^2{d\tau\over d M^2}.
\end{equation}
Near $\tau\approx 0$, where $u\approx +\Lambda^2$, the dual Higgs
VEV, $a_D$, goes like $a_D\propto (u-\Lambda^2)/\Lambda$
and the monopole mass is $M_D=\sqrt{2} a_D$ so 
$\Sigma_{+1}$ gives
\begin{equation}
\Sigma_{+1}\approx -(u-\Lambda^2){d\tau\over du}\approx -a_D{d\tau\over da_D}
\approx -{M^2_D\over 2}{d\tau\over dM^2_D}.
\end{equation}
$\Sigma_{+1}$ reduces to the Callan-Symanzik $\beta$-function close to the two dual
points $\tau=i\infty$ and $\tau=0$, apart from an overall factor of $2$ at $\tau=0$.
Away from these points its physical interpretation is not so clear.

There is however still a pathology in $\Sigma_{+1}$ since
there is singularity at $u=-\Lambda^2$ where $\tau=1$.
This can be avoided by using $\Sigma_{-1}$
to describe the crossover from $u=-\infty$ to $u=-\Lambda^2$
and the discussion exactly parallels that for $\Sigma_{+1}$ except
that $\tau$ is replaced with $\tau+1$ or equivalently
$\vartheta_3$ is replaced with $\vartheta_4$.
The flow is that of figure 1 with the horizontal axis displaced
by one unit in either direction.
Any starting value $0<\theta<4\pi $ at weak coupling
is driven to $2\pi$
at strong coupling.
There is no holomorphic scaling function compatible with $\Gamma(2)$
symmetry which has no singularities at all, since any modular function of
weight -2 must have at least one singularity somewhere within, or on the
boundary of, the fundamental domain --- at best the scaling function
is meromorphic.

In fact one can do better and define a scaling function that
gives the correct behaviour at all three singular points in the $u$-plane
(from now on we shall set $\Lambda=1$ so these singularities are at
$u=\infty$, $u=+1$ and $u=-1$).  We want to construct a scaling function
that is meromorphic in $u$
and vanishes at both $u=\pm 1$, without disturbing the behaviour 
$\Sigma\approx -{i\over\pi}$ at $u\approx i\infty$ and this can 
be done using the ideas of Ritz in \cite{RDBMLL}. 
If we wish to avoid extraneous poles or zeros the only
possibility is 
\begin{equation}
\Sigma=-\left({(u-1)^m(u+1)^n\over u^{m+n}}\right) u{d\tau\over du}
\end{equation}
with $m$ and $n$ positive integers. 

Demanding the correct asymptotic behaviour at $u\approx \pm 1$ requires
$m=n=1$ so
\begin{equation}
\Sigma(\tau)=
-{(u^2-1)\over u}{d\tau\over du}=
{2\over \pi i}{1\over \big(\vartheta_3^4(\tau) + \vartheta_4^4(\tau)\bigr)}
\quad \mathop{\longrightarrow}_{\tau\rightarrow i\infty}\quad
{1\over \pi i}.
\label{betaG02}
\end{equation}
This flow is shown in figure 3. The behaviour near $\tau=0$ is
\begin{equation}
\Sigma(\tau)\approx {2i\over \pi}\tau^2
\end{equation}
or, in terms of the dual coupling $\tau_D$,
\begin{equation}
\Sigma(\tau_D)\rightarrow {2i\over \pi}
\end{equation}
as $\tau_D\rightarrow i\infty$.
This is twice $\Sigma_{+1}(\tau_D)$ as $\tau_D\rightarrow i\infty$,
because of the pre-factor $u+1\rightarrow 2$ as $u\rightarrow 1$.
It is therefore a factor of 4 greater than Callan-Symanzik 
$\beta$-function at this fixed point.
 
By construction (\ref{betaG02}) is symmetric under $\tau\rightarrow
\tau +1$ and so is a modular function for $\Gamma_0(2)$, just
as $\Sigma_0$ was,  but now there is a singularity at $\tau={1+i\over 2}$
(where $\vartheta_3^4=-\vartheta_4^4$)
corresponding to a repulsive fixed point, rather than
the attractive fixed point of $\Sigma_0$.

There is an infinite hierarchy of crossovers near the real axis in figure 3
as a consequence of $\Gamma_0(2)$ symmetry.  Under the action of an
element $\gamma=\pmatrix{a & b\cr c & d\cr}$ of $\Gamma_0(2)$, 
$\tau\rightarrow {a\tau +b\over c\tau +d}$ (with 
$c$ even and $ad-bc=1$) the point $\tau=i\infty$
is mapped to $\gamma(\tau)=a/c$, $\tau=0$
to $\gamma(\tau)=b/d$ and $\tau=1$ to $\gamma(\tau)={(a+b)/(c+d)}$.
As observed in \cite{SW}, the infinite hierarchy of vacua at 
strong coupling can thus be classified into
three types, in terms of $\Gamma_0(2)$ these are: 
images of $\tau=i\infty$ with $\theta/2\pi=a/c$, a rational number
with even denominator; 
images of $\tau=0$ with $\theta/2\pi=b/d$, a rational number 
with odd denominator and
images of $\tau=0$ with $\theta/2\pi=(a+b)/(c+d)$ again a rational number
with odd denominator.
The fermionic degrees of freedom at $\tau=i\infty$ are gluinos with 
electric charge $+1$
and magnetic charge $0$, at $\tau=0$ they are monopoles with electric charge
zero and magnetic charge $+1$ while at $\tau=1$ they are dyons with
electric charge $-1$ and magnetic charge $+1$.  In general, at a strong
coupling fixed point,
$\theta/2\pi=-q/m$ where $q$ is the electric charge and $m$ is 
the magnetic charge of the massless fermionic degrees of freedom,
which are composite objects in terms of
the relevant degrees of freedom at weak coupling.
  
Let us now look a little more closely at the flow structure
generated by $\Sigma$
near the real axis in figure 3. 
As $u$ varies between $+\infty$ and $+1$,
$\Sigma$ generates a semi-circle between
two states on the real line,
$\theta_1/2\pi=-q_1/m_1$ and $\theta_2/2\pi=-q_2/m_2$ 
(we shall assume that $q_1$ and
$m_1$ are mutually prime, similarly for $q_2$ and $m_2$). 
This semi-circle can be obtained
from the the positive imaginary axis in the $\tau$-plane,
with end points $i\infty$ and $0$,
by the action of some $\gamma=\pmatrix{a&b\cr c&d\cr}\in\Gamma_0(2)$, 
with $a$ and $d$ odd and $c$ even.  Thus $q_1/m_1=-a/c$ and
$q_2/m_2=-b/d$, so $q_1=\pm a$ is odd and $m_1= \mp c$ even,
$q_2=\pm b$ is of undetermined parity while $m_2=\mp d$ is odd.  
Since $ad-bc=1$ we see that 
\begin{equation}
q_1m_2-q_2m_1=\pm 1
\label{selectionrule}
\end{equation}
and we have
a selection rule for transitions between 
vacua in the strong coupling regime as $u$ is varied.  

The flow from $u=-1$ to $u=+1$, which is the
semi-circular arch spanning $\tau=1$ to $\tau=0$ in the $\tau$-plane, 
cannot be obtained from the flow along the imaginary axis in the
$\tau$-plane by using $\Gamma_0(2)$ so we consider this separately.
This semi-circle is mapped to another semi-circle 
linking $(a+b)/(c+d)$ to $b/d$:
that is linking $\theta_1/2\pi=-q_1/m_1=(a+b)/(c+d)$ 
(so $m_1$ is odd) to 
$\theta_2/2\pi=-q_2/m_2=b/d$ (so $m_2$ is odd).  
Again $q_1m_2-q_2m_1=\pm 1$.

This selection rule is related to the Schwinger-Zwanziger quantisation
rule for a pair of Dyons with charges $(Q_1,M_1)$ and $(Q_2,M_2)$
\cite{SZ}:
\begin{equation}
Q_1M_2-Q_2M_1=4\pi n
\end{equation}
with $n$ an integer.\footnote{More generally $n$ could 
be half-integral, but in pure SUSY Yang-Mills we
are always dealing with $U(1)$ charges coming from the adjoint representation
of $SU(2)$ so it is integral in this case.}
Writing $Q_i=q_ig$ and $M_i=m_i{4\pi\over g}$, with $q_i$ and $m_i$
integers, this is 
\begin{equation}
q_1m_2-q_2m_1=n
\end{equation}
and the selection rule (\ref{selectionrule}) dictates that 
flow always connects the nearest pairs allowed
by the Schwinger-Zwanziger quantisation rule.

\section{The Quantum Hall Effect}

The interpretation of the infinite hierarchy of states 
for N=2 SUSY Yang-Mills presented
in the previous section 
is very similar to the hierarchy of states observed in the
quantum Hall effect (a connection between $N=2$ SUSY Yang-Mills and
the quantum Hall effect was suggested in \cite{Crossover}).  
For the quantum Hall effect the complex coupling $\tau$ is replaced
with a complex conductivity, $\sigma=\sigma_{xy}+i\sigma_{xx}$,
with $\sigma_{xy}$ the Hall conductivity and $\sigma_{xx}$ the
Ohmic conductivity (for an isotropic layer with $\sigma_{xx}=\sigma_{yy}$).
At Hall plateaux (where $\sigma_{xx}=0$) the Hall conductivity 
is quantised,
in units in which $e^2/h=1$, as a rational number $\sigma_{xy}=q/m$
where $m$ is odd.  These plateau are attractive fixed points of a 
scaling flow, even denominators being repulsive, \cite{KH, LLP}.
The suggestion in \cite{ShapereWilczek} that the modular
group might be relevant to the
quantum Hall effect was further developed in 
\cite{LR1, Crossover, BL, LR2,FK,GWBD,Duality}.
Indeed the group $\Gamma_0(2)$ has an action 
known as the \lq \lq Law of Corresponding States'' in the condensed
matter literature \cite{KLZ,JKN}, which holds under the
assumption of well separated Landau levels with completely
spin polarised electrons (when spin effects are important
it was argued in \cite{Gamma2} that the $\Gamma_0(2)$ symmetry is broken
to $\Gamma(2)$).

Constraints on the scaling functions for the quantum Hall effect,
as a result of modular symmetry, 
were discussed in \cite{BL}.  If meromorphicity is assumed
stronger statements can be made and a meromorphic flow 
diagram was presented 
in \cite{Crossover}, 
which developed the original
flow suggested in \cite{KH}.  This meromorphic flow is given 
by\footnote{Up to an undetermined
constant which is chosen to agree with (\ref{betaG02}) here.}
\begin{equation}
\Sigma(\sigma)={2\over\pi i}{1\over \bigl(\vartheta_3^4(\sigma) + 
\vartheta_4^4(\sigma)\bigr)}
\label{QHEflow}
\end{equation}
and is identical to figure 3 though the physical interpretation is
different. For the quantum Hall effect the flow lines are in the direction
of increasing effective system size (in practice this can be translated
to decreasing temperature) \cite{Huckestein} 
and different lines correspond to
different values of the external magnetic field.
As the temperature is lowered and the flow runs down from 
large $\sigma_{xx}$, $\sigma$ is driven onto a semi-circular arch
in the complex $\sigma$-plane, such as the semi-circle connecting
$\sigma=2$ to $\sigma=1$ in figure 3 for example.  
On this semi-circle, there is
a second order quantum phase transition,
as $T\rightarrow 0$,  at $\sigma=(3+i)/2$.  At or close to $T=0$,
$\sigma$ becomes a function of a single scaling variable, 
$\sigma(\Delta B/T^\mu)$, where $\Delta B = B_c$ is the deviation
of the magnetic field from its critical value and $\mu$ a scaling
exponent for the temperature \cite{KH,LLP, Huckestein}.  
In this regime of very low temperatures $\sigma_{xy}$ runs
between $2$ and $1$ as $B$ is varied at fixed $T$.  

The picture is the same at all copies of the semi-circle under the
action of $\Gamma_0(2)$, the exponent $\mu$ is believed to be
the same for every transition, 
a phenomenon known as \lq super-universality',
and there is good experimental evidence that this is indeed the case \cite{SU}
(the experimental situation is a little murky on this point
however, in some experiments scaling appears to be  violated
at low temperatures \cite{SHLTSR}).
Under the assumption of $\Gamma_0(2)$ symmetry 
the critical points above the real axis
are fixed points of  $\Gamma_0(2)$\footnote{i.e. there 
exists an element of $\Gamma_0(2)$
which leaves the point invariant.}
and so their positions
can easily be calculated for the crossover between any two
given plateaux.

The derivation of the $\Gamma_0(2)$ flow in \cite{Crossover}
was made under the following assumptions:
i) in the long wavelength limit
the flow should commute with the action of $\Gamma_0(2)$,
in particular this implies that any point which is a fixed point
of $\Gamma_0(2)$
should also be a fixed point of the
flow; ii) there are no fixed points of the flow that are not fixed
points of $\Gamma_0(2)$; iii) the scaling functions are modular
forms of weight -2, in the sense of \cite{Koblitz}
(in particular they are meromorphic);
iv) due to the stability of the quantum Hall plateaux, the flow
should approach the real axis at rational numbers with odd denominators 
(attractive fixed points of the flow) as fast as possible;
v) the flow should be vertically downwards when the Ohmic conductivity
is large.  

Assumption iii) is an assumption about
the analyticity properties of flow.  
Mapping $\sigma \rightarrow \gamma (\sigma)=(a\sigma +b)/(c\sigma +d)$ 
we have, under any variation $\delta\sigma$ of $\sigma$,
\begin{equation}
\delta(\gamma(\sigma)) = {1\over (c\sigma +d)^2} \delta\sigma
\end{equation}
since $ad-bc=1$, so $\Sigma(\sigma)$ is automatically a modular function
of weight $-2$ if it is meromorphic.
The general form of the flow does not depend crucially on 
assumption iii), provided i) and ii) hold 
small deformations away from meromorphicity cannot change
the topology of the flow or even the position of the fixed points ---
any such deformation would smoothly distort the lines of figure 3 but 
must leave the
fixed points  and the topology invariant.  Experimental
flow diagrams are in remarkable agreement with the $\Gamma_0(2)$ predictions:
the flow diagram in \cite{Murzin1}
found for the integer quantum Hall effect is reproduced in figure 4
and that found in \cite{Murzin2} for the fractional effect in figure 5.

Another consequence of $\Gamma_0(2)$ symmetry for the quantum Hall effect
is the semi-circle law.  
Semi-circles in the upper-half complex plane,
spanning certain pairs of rational points on the real axis,
are mapped into one another by the action of the modular group and so are 
rather special curves.
Experimentally the crossover between two plateaux, as the external
magnetic field is varied at fixed low temperature, is often
very close to a semi-circle \cite{Hilke}.  
In the condensed matter literature this is known
as the \lq \lq semi-circle law'' and it can be interpreted as a consequence
of $\Gamma_0(2)$ symmetry.  Even
without assuming any  meromorphicity properties, just assuming that
$\Gamma_0(2)$ commutes with the flow and there is a symmetry between
the pseudo-particles and the accompanying holes, the
semi-circle law for the crossover between quantum Hall
plateaux can be derived \cite{semicircle}.

$\Gamma_0(2)$ symmetry also implies 
a selection rule $q_1m_2-q_2m_1=\pm 1$ for transitions
between quantum Hall plateaux as the magnetic field is varied, 
\cite{SelectionRule}.  This rule is only expected to hold
for two well formed plateaux with no hint of unresolved sub-structure 
between and is very well supported by experimental data on the
quantum Hall effect when this is the case, at least for quantum Hall
monolayers with the spins well split.\footnote{In figure 4,
which is interpreted as a transition from $1/1$ to $0/1$, the spins
are degenerate, which doubles the degrees of freedom in the 
lowest Landau level and so doubles the conductivity \cite{Murzin1}.}

The selection rule and the semi-circle law do not require any
assumptions about the analytic properties of the flow --- they
only require that the action of $\Gamma_0(2)$ commutes with it.

Another similarity between $N=2$ SUSY Yang-Mills and the quantum Hall effect 
is that,
in the composite boson picture of the quantum Hall effect described in
\cite{KLZ},
the effective degrees of freedom 
in a state with $\sigma_{xy}=1/m$ are electrons
with $m$-units of magnetic flux attached with $m$ odd, that is
the quasi-particles are composite objects with an odd number of
vortices attached to fundamental charged particles.

While, as a mathematical theory, $N=2$ SUSY has an infinite hierarchy
of vacuum phases, the quantum Hall effect, as a physical phenomenon,
does not.  Experimentally there are limitations to the applicability
of $\Gamma_0(2)$ in any given sample and not all mathematically allowed
phases will be seen.   For very strong magnetic fields, with filling factors
less than about $1/7$,
it is believed that the 2-dimensional
electron gas will enter a new phase at $T=0$, the Wigner crystal, which is not
part of the modular hierarchy, though factors as low as $1/9$
have been seen at finite temperature \cite{PSTPBW}.  In addition only
fractions up to a maximum denominator, 
depending on the sample, will be seen ---
other physical factors, such as impurities, limit how far into the hierarchy 
one can penetrate.  Also if the temperature is too high the quantum
Hall effect is destroyed so there is a limit as to how high
up into the complex conductivity plane one can trust the flow derived
by assuming $\Gamma_0(2)$ symmetry.

\section{Conclusions}

It has been argued that scaling functions can be defined
for $N=2$ supersymmetric Yang-Mills, without matter fields, which
are modular forms of $\Gamma_0(2)$ and which reduce to the
correct Callan-Symanzik $\beta$-functions , both at strong
and at weak coupling, up to a constant.  The flow is shown in figure 3 and
there is one singular fixed point 
in the fundamental domain, at $u=0$,
where the classical theory would have full $SU(2)$ gauge symmetry
restored but the $W^\pm$ bosons remain massive in the quantum theory. 
Not only does the quantum theory prevent the classical symmetry restoration,
it actively avoids the point $u=0$ in both flow directions.
Massless dyons in the strong coupling regime have electric charge $q$ 
and magnetic charge $m$ with $m$ odd for attractive fixed points
and $m$ even for repulsive fixed points (in the flow direction in which
the Higgs mass is lowered) and $\theta=-q/m$ is a rational fraction
as $g\rightarrow\infty$. 

Exactly the same flow pattern has been predicted theoretically
for the long wavelength physics of the quantum Hall effect,
on the basis of $\Gamma_0(2)$ symmetry 
(the \lq \lq Law of Corresponding States'') 
and observed experimentally.  In the quantum Hall effect the
parameter governing the flow is temperature and odd denominator plateaux 
are attractive fixed points while even denominator plateaux are 
repulsive.\footnote{There are exceptions to this at high Landau levels,
the most famous being the $5/2$ state.  These states can be interpreted
as being a due to Bose-Einstein condensation of pairs of composite
Fermions and are not part of the $\Gamma_0(2)$ hierarchy.}
Attractive fixed points give rise to quantised Hall plateaux where
the Hall conductivity $\sigma_{xy}=q/m$ is rational with odd denominator.

In both cases the effective degrees of freedom are composite objects
in terms of the fundamental degrees of freedom --- dyons in the case
of SUSY Yang-Mills and composite fermions or composite bosons in the 
case of the quantum Hall effect.  Composite bosons (odd denominator states) 
are charged particles carrying an odd number of magnetic vortices \cite{KLZ}
and composite fermions (even denominator states) carry an even number
of magnetic vortices \cite{Jain} (in both cases the underlying particles
are still fermionic).

It would seem that $N=2$ supersymmetric Yang-Mills in $3+1$
dimensions has much in common with the quantum Hall effect in $2+1$
dimensions but exactly what the relation is, what are the essential
features that are required to bring out modular symmetry in the effective
action, is unclear at this stage,
and there is scope for much work in the future to clarify
this relation (the modular group was also found to play a role in 
in other models \cite{Cardy,Callan} and was considered in
$2+1$ dimensional abelian Chern-Simons theory in \cite{Witten}).
There is no suggestion here that supersymmetry is relevant to
the quantum Hall effect, rather it would appear that some minimal
set of criteria is necessary for modular symmetry to emerge at long
wavelengths (general criteria for any system 
in $2+1$ dimensions were discussed in \cite{Duality}).  
Supersymmetry is not essential, since nature has given us
an experimental system which exhibits modular symmetry without it, 
and neither is Lorentz invariance.  The number of dimensions can be
either 3 or 4 (from the field theory point of view the quantum Hall
effect is a 3-dimensional phenomenon, since it occurs at very low
temperatures).  Perhaps other dimensions are possible too.  
Certainly topological effects are essential and effective degrees
of freedom that are composite objects in terms the fundamental
degrees of freedom and the topologically non-trivial degrees of
freedom are common to both systems.

It is a pleasure to thank the Mathematics Department, Heriot-Watt 
University, Edinburgh, Scotland, where part of this work was
undertaken, as well as the Perimeter Institute, Waterloo,
Canada, for hospitality.  The work was partially funded by
a UK Royal Society short term visit grant.

\newpage

\vtop{
\hbox{\epsfxsize=15cm
\hskip -25pt \epsffile{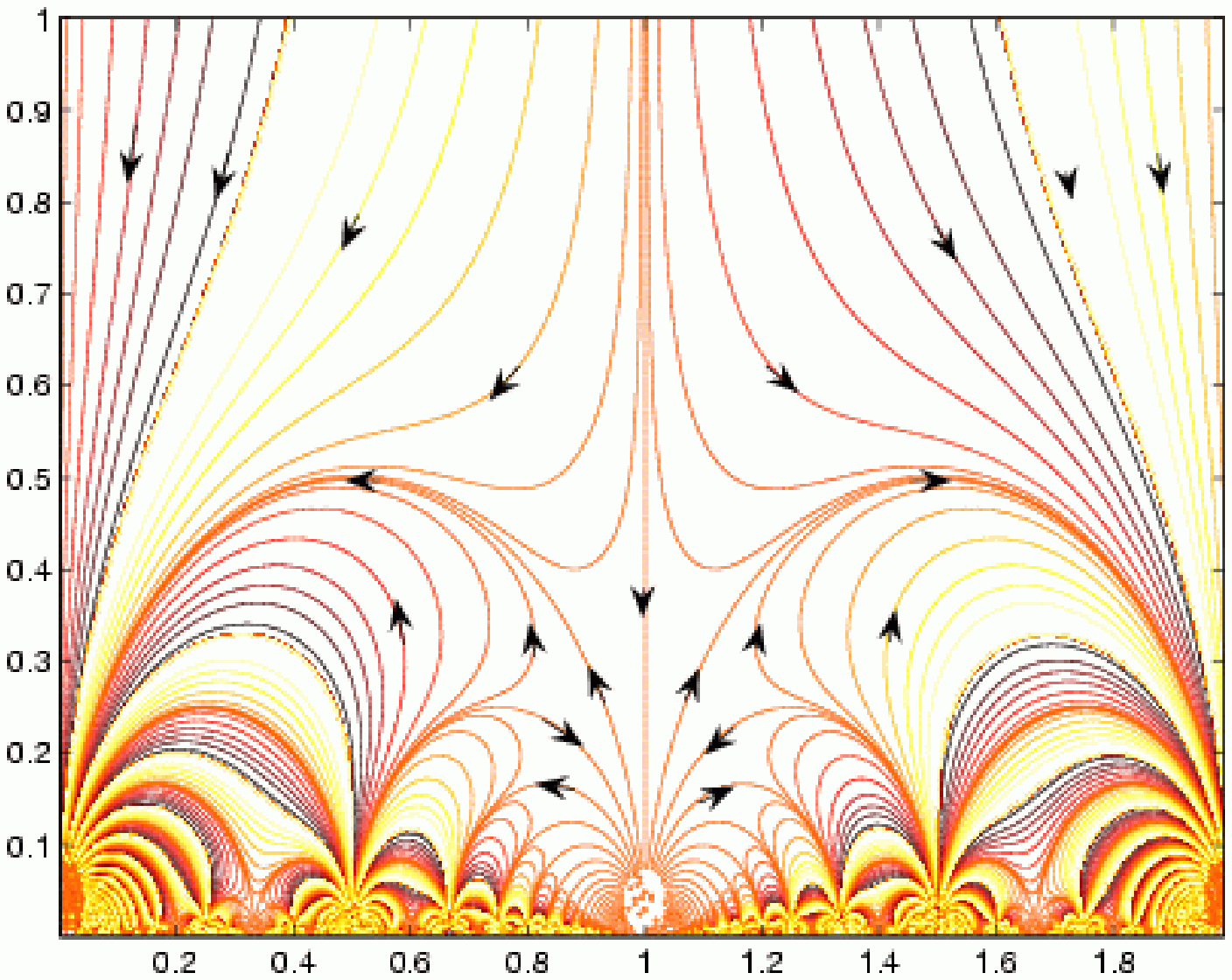}}
\noindent Fig. 1: Flow of effective coupling  
of $N=2$ SUSY Yang-Mills, as the Higgs VEV is reduced, using 
(\ref{betaone}). The pattern repeats under $\tau\rightarrow \tau+2$.
There is a singularity at $\tau=1$ and its images under $\Gamma(2)$. 
Note that 
$\theta$ is driven to zero at strong coupling for any
starting value between $-2\pi$ and $2\pi$ at weak coupling. }

\newpage

\vtop{\epsfxsize=10cm
\hbox{\hskip 50pt \epsffile{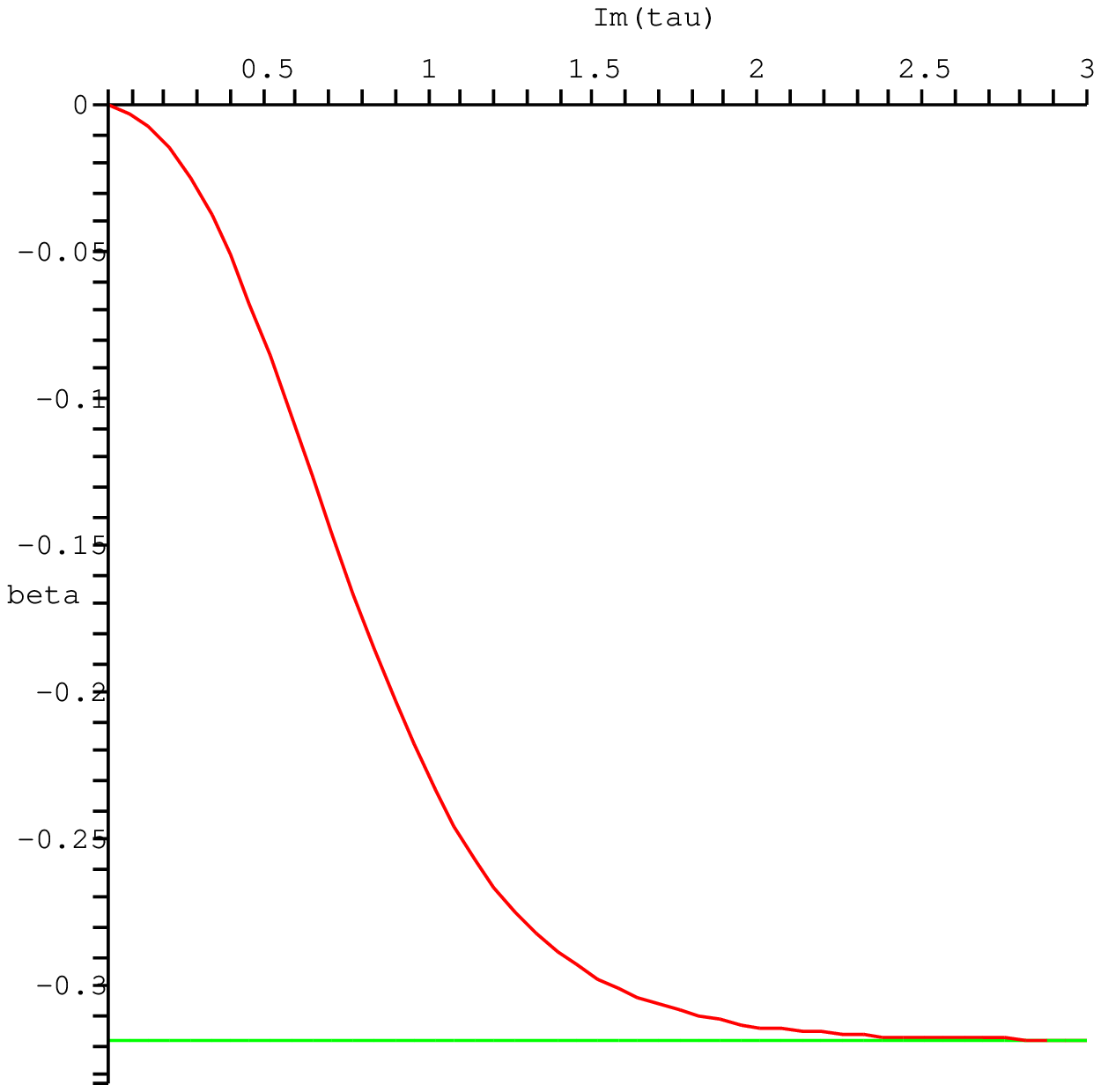}}
\noindent Fig. 2: Crossover of $\hbox{Im}\tau$ from weak to strong coupling
along the imaginary axis $\theta=0$.
}

\newpage

\vtop{\epsfxsize=15cm \epsfysize=10cm
\hbox{\hskip -20pt \epsffile{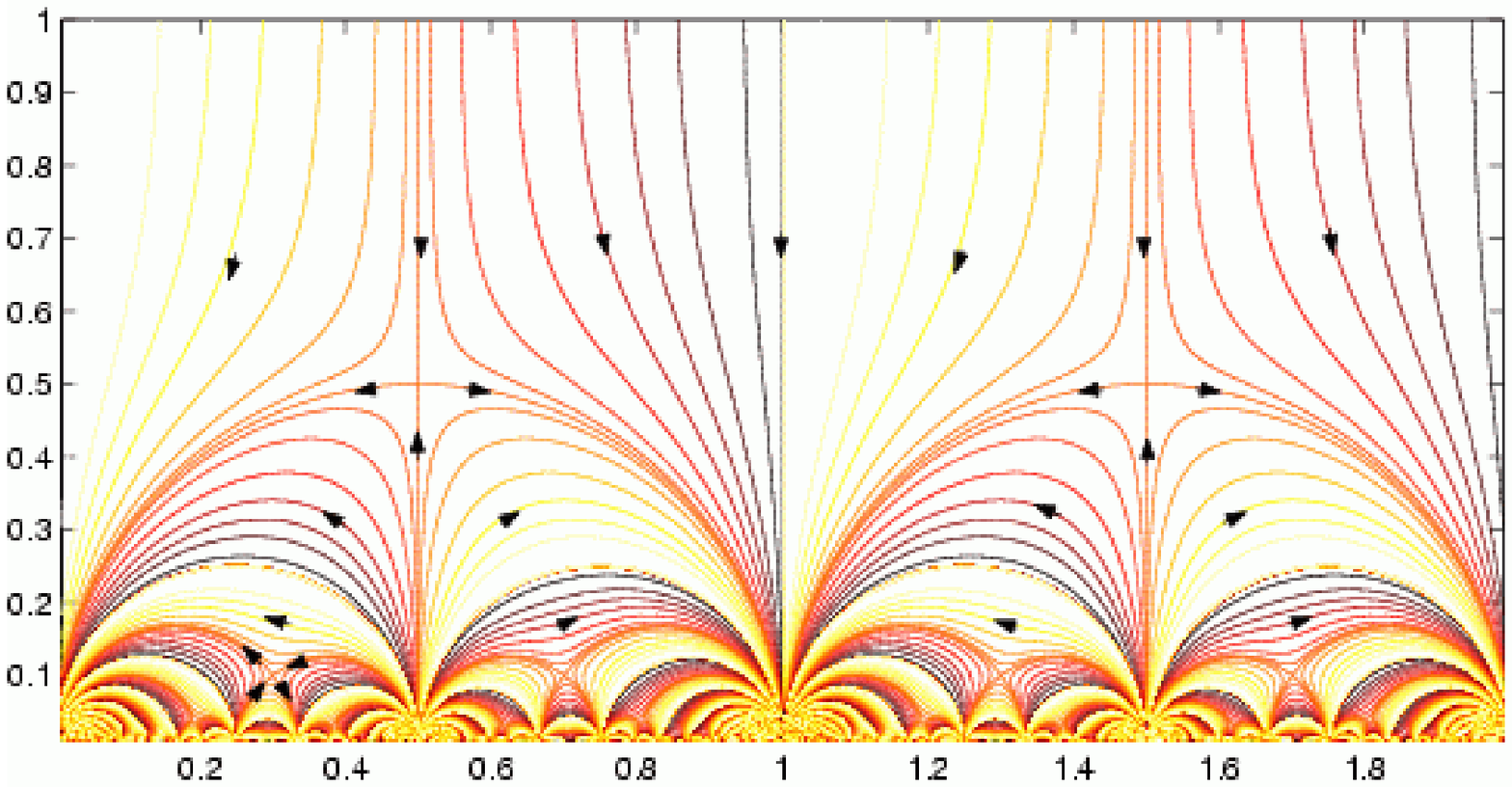}}
\noindent Fig. 3: Flow of effective coupling for $N=2$ SUSY 
Yang-Mills, as the
Higgs VEV is reduced, using (\ref{betaG02}).  The same flow applies
to the complex conductivity of the quantum Hall effect as the
temperature is lowered.  
The pattern repeats under $\tau\rightarrow\tau+1$.
Note the repulsive fixed points at $(1+i)/2$ and its images under
$\Gamma_0(2)$.}

\newpage

\vtop{\epsfxsize=10cm
\hskip 10pt \hbox{\epsffile{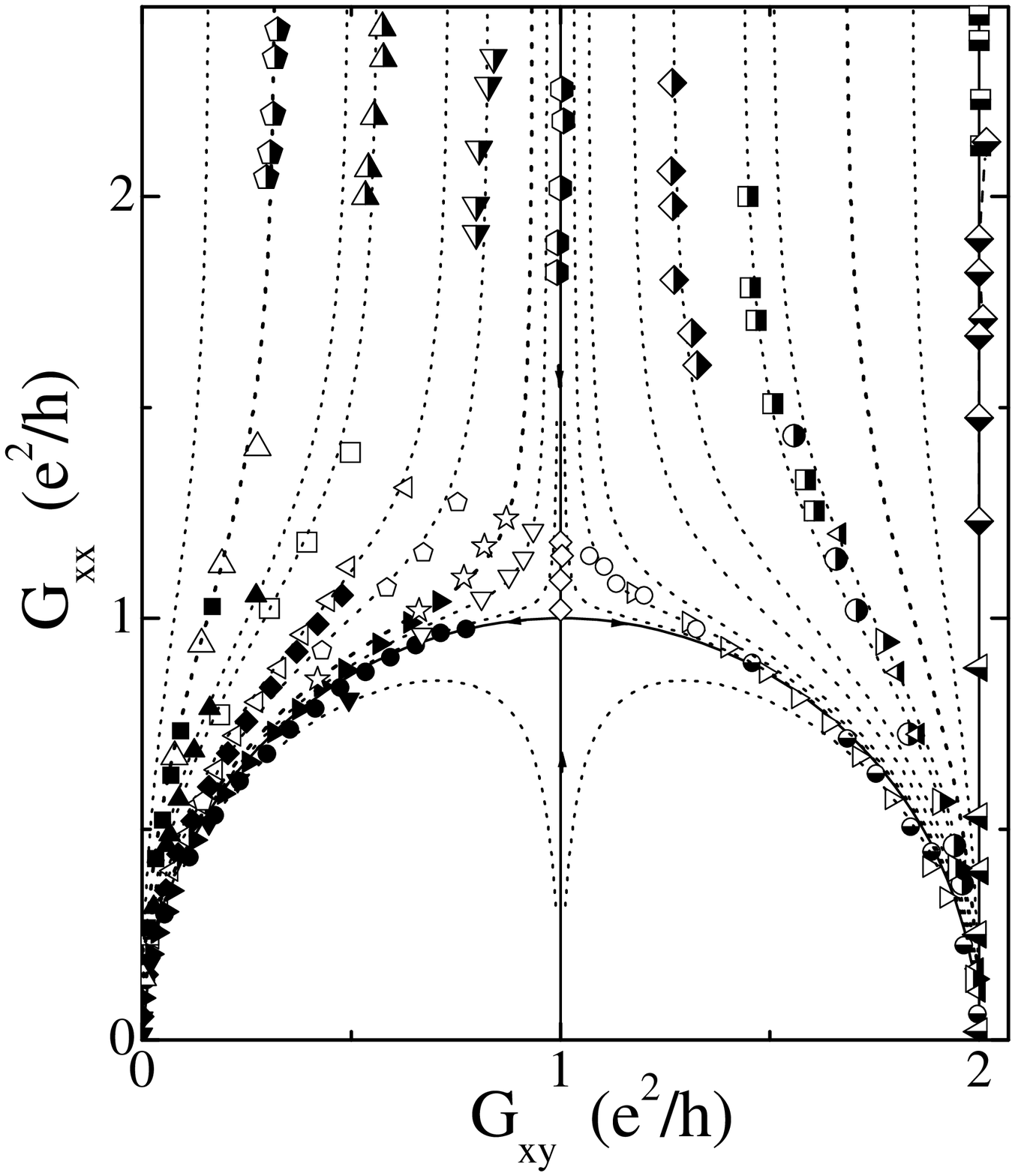}}

\noindent Fig. 4: As the temperature is lowered the conductivity 
flows down from $\sigma_{xx}=\infty$, different flow lines correspond
to different magnetic fields.  The dotted lines are the
flow following from $\Gamma_0(2)$ symmetry and meromorphicity
(equation (\ref{QHEflow})) and the symbols are experimental data.
(The conductivities are twice those in the
text due to spin degeneracy --- the Landau levels in the sample used
here are spin degenerate, so conductivities are multiplied by 2 and
$\Gamma_{0}(2)$ acts on $\sigma/2$ rather than on $\sigma$.)
Figure reproduced from \cite{Murzin1}.}

\newpage

\vtop{\epsfxsize=17cm

\hskip -60pt \hbox{\epsffile{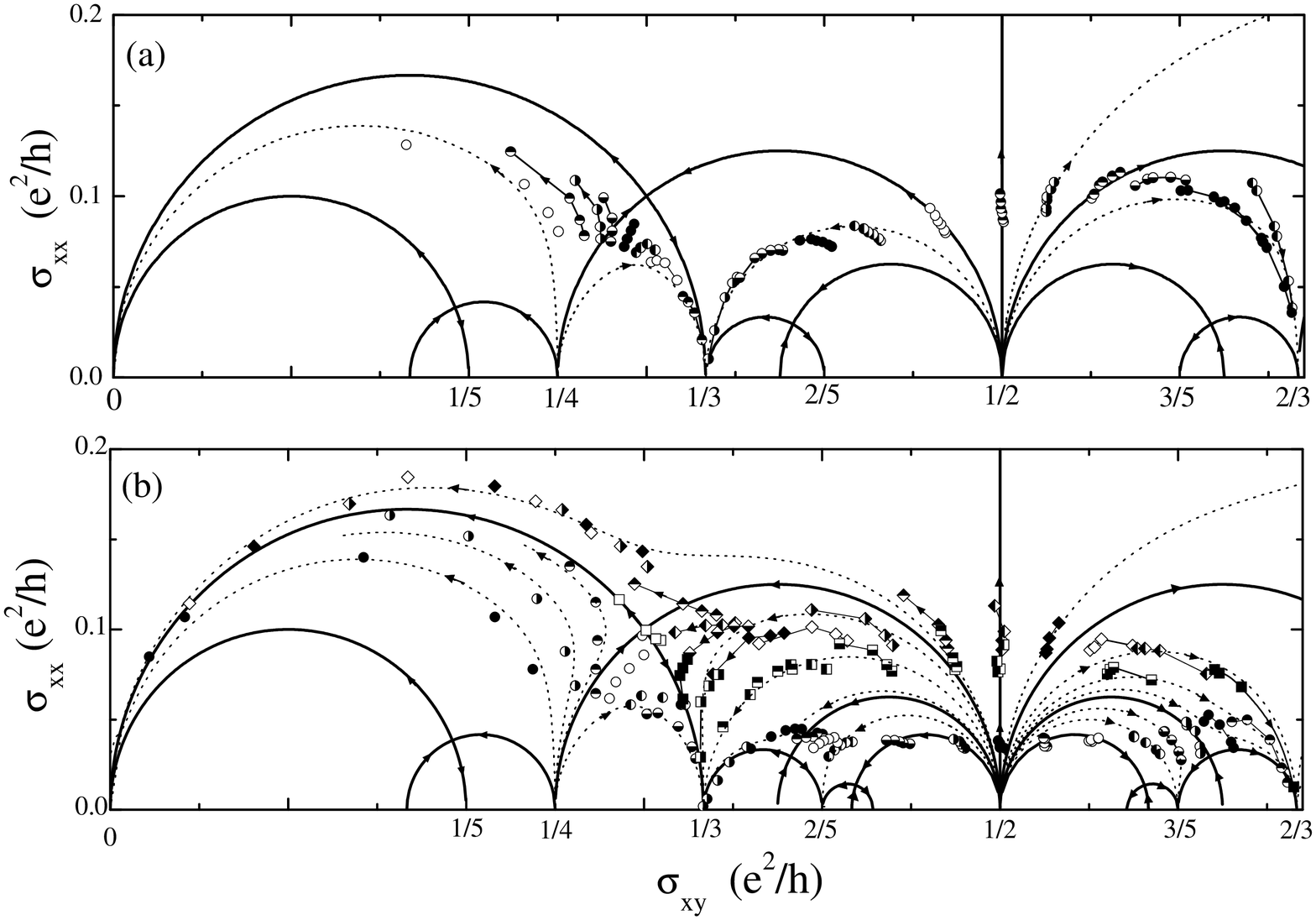}}

\noindent Fig. 5: Temperature flow for the fractional quantum Hall effect.
The upper and lower figures, (a) and (b), represent two different samples.
Dashed and solid lines are the
flow following from $\Gamma_0(2)$ symmetry and meromorphicity
(equation (\ref{QHEflow})) and the symbols are experimental data.
Note the repulsive fixed point at $\sigma=1/2$ (figures taken from
\cite{Murzin2}).}

%\newpage

%\vtop{\epsfxsize=15cm
%\hbox{\epsffile{QHE-flow.eps}}
%
%\hskip 7cm \hbox{Fig. 3}}

\end{document}